\documentclass[prl,twocolumn,showpacs,superscriptaddress,nofootinbib]{revtex4-1}
\usepackage{graphicx}
\usepackage{bm}
\usepackage{amssymb}
\usepackage{color}
\usepackage[usenames,dvipsnames]{xcolor}
\usepackage{amsmath}
\usepackage{amstext}
\usepackage{latexsym}
\usepackage[colorlinks=true,citecolor=Cerulean,linkcolor=RubineRed,urlcolor=ForestGreen]{hyperref}

\def\q{{\rm\bf q}}

\def\la{\langle}
\def\ra{\rangle}

\newcommand{\beq}{\begin{equation}}
\newcommand{\eeq}{\end{equation}}
\newcommand{\beqa}{\begin{eqnarray}}
\newcommand{\eeqa}{\end{eqnarray}}

\newcommand{\nab}{{\bf{\nabla}}}

\newcommand{\pa}{\partial}

\newcommand{\hanaV}{\mathcal{V}}

\begin{document}

\title{High-fidelity rapid ground-state loading of an
ultracold gas into an optical lattice}
\author{Shumpei Masuda}
\affiliation{The James Franck Institute, The University of Chicago, Chicago, 
Illinois 60637, USA}
\affiliation{Department of Physics, Tohoku University, Sendai
980, Japan}
\author{Katsuhiro Nakamura}
\affiliation{Turin Polytechnic University in Tashkent, 17 Niyazov Street, Tashkent 100095, Uzbekistan}
\affiliation{Department of Applied Physics, Osaka City University, Sumiyoshi-ku, Osaka 558-8585, Japan}
\author{Adolfo del Campo}
\affiliation{Theoretical Division, Los Alamos National Laboratory, 
Los Alamos, NM 87545, USA}
\affiliation{Center for Nonlinear Studies, Los Alamos National Laboratory, 
Los Alamos, NM 87545, USA}

\date{\today}
\begin{abstract}
A protocol is proposed for the rapid coherent loading of 
a Bose-Einstein condensate into the ground state of an optical lattice, 
without residual excitation associated with the breakdown of adiabaticity.
The driving potential required to assist the rapid loading is derived using the 
fast forward technique, and
generates the ground state in any desired short time.
We propose an experimentally feasible loading
scheme using a bichromatic lattice potential, which approximates 
the fast-forward driving potential with high fidelity.
\end{abstract}
\maketitle


 The advance of quantum technologies based on ultracold atoms 
requires new control methods to manipulate matter waves towards the 
preparation of a given
target state. 
Coherent controls of Bose-Einstein condensates (BECs) 
and molecular chemical reactions have
been demonstrated experimentally \cite{shin,ric2}.
Most of the control schemes 
resort to adiabatic driving 
and their efficiency is limited by a variety of uncontrolled effects
including decoherence, three-body losses, 
noise sources, etc. 
 This motivates the growing body of experimental
and theoretical work devoted to tailor excitations in nonadiabatic 
processes and hence, to design of shortcuts to
adiabaticity (STA) \cite{Torrontegui13}.
 It is now a well-established fact
that experimentally realizable STA can be found for the
fast driving of matter-waves \cite{Chen10,Hoff,Jarz,Stef}, 
BECs 
\cite{Muga10,delcampo11epl,Stef2} and  other many-body systems \cite{delcampo11,DB12,Hoff2,delcampo13,NEWREF_A} in self-similar processes,
see \cite{Deff} for a unifying framework. 
Experimental demonstrations have been reported with thermal clouds \cite{Schaff1}, BECs \cite{Schaff2,Bas} and a single trapped ion \cite{ions1,ions2}.
A particularly remarkable achievement is the recent realization of
STA at the many-body level in tightly confined ultracold gases \cite{Rohr}.
The extension of these techniques to processes not governed by scaling laws remains challenging.
Designing an experimentally realizable protocol generally demands the knowledge of the spectral 
properties of the system \cite{DR03,Berry09} and the
resulting counterdiabatic fields might involve non-local
interactions \cite{DMZ12,delcampo13}. 
The fast-forward approach developed by Masuda and Nakamura \cite{mas1,mas2,mas3,Torrontegui12,mas4} provides a remarkable way out of this problem.
Indeed, it has proved useful in the design of realizable schemes for the fast driving of dynamical processes lacking self-similarity, e.g. splitting of matter waves \cite{Torrontegui13b}. 
\begin{figure}[t]
\begin{center}
\includegraphics[width=5.5cm]{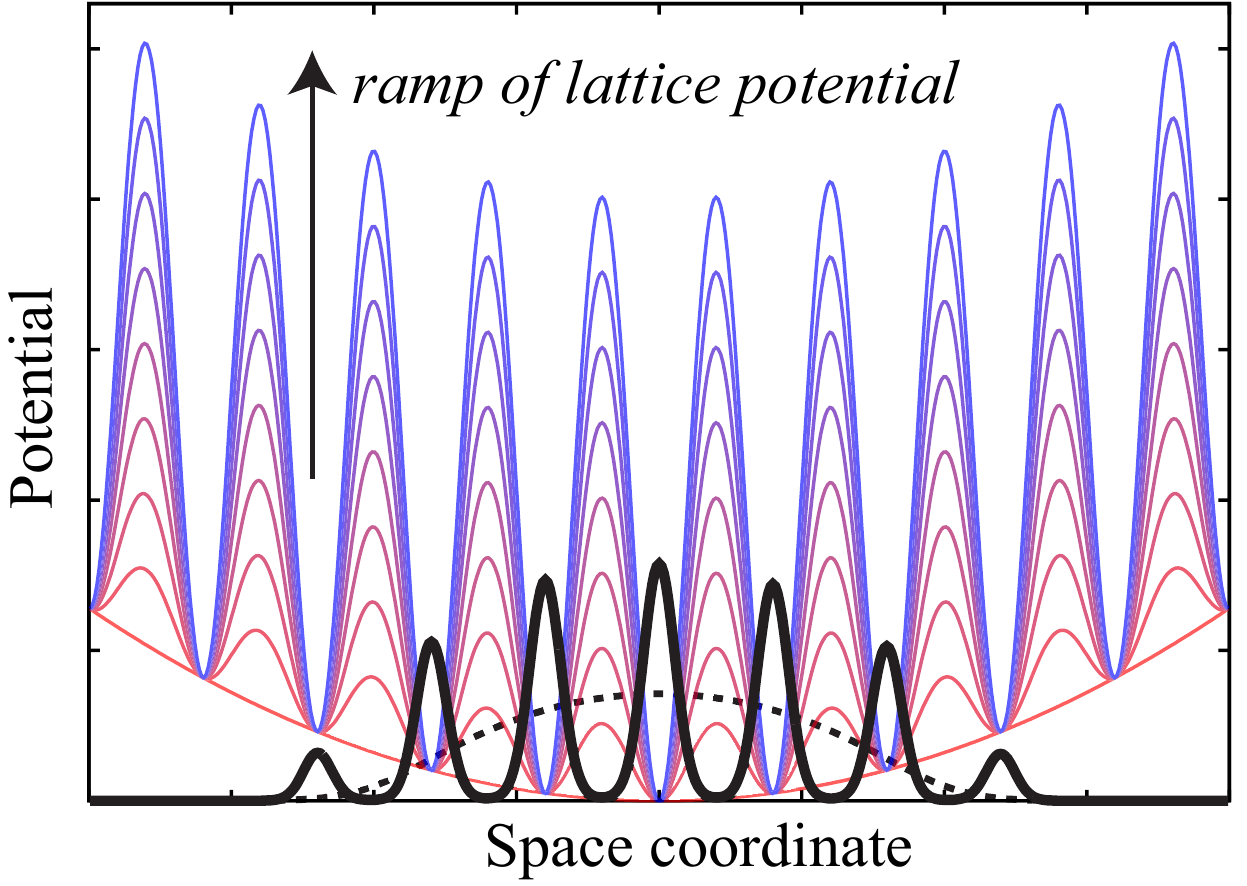}
\end{center}
\caption{\label{fig:epsart} (Color online) Schematic diagram of the adiabatic 
loading of a BEC into the ground state of an optical  lattice potential.
A dotted curve and a thick solid curve represent the ground state in a
harmonic trap and the one in a harmonic and an optical lattice traps, 
respectively.
Thin solid curves represent the adiabatic ramp of the lattice potential.
}
\label{potential2}
\end{figure}

These more general processes are ubiquitous in the study of ultracold gases
in optical lattices, whose research is spurred by  many applications including, 
nonlinear physics \cite{MO06},
quantum simulation \cite{BDZ08,LSA},
 the realization of optical-lattice atomic clocks \cite{DK11}, 
and even quantum information processors, which requires 
precise controls of quantum states 
and to reduce the effect of decoherence \cite{LSA,JZ05}.
A recurrent goal is to prepare the ground state of the system, e.g., a Bose-Einstein condensate (BEC), in 
the optical lattice without residual excitations.
A natural strategy is to start  with an atomic cloud confined in a shallow harmonic trap, 
and ramp adiabatically the sinusoidal optical-lattice potential 
created by laser fields  (see Fig. \ref{potential2}).
Applications  requiring the coherence of the cloud and the concatenation of
many operations during the coherence time call for alternative control techniques \cite{CZ12,LSA,JZ05}. Schemes for rapid loading have been studied optimizing  parameters
such as the duration of the ramp time and the intensity of the laser fields \cite{Heck,Mel,Jul,Liu}, including self-similar STA \cite{NEWREF_B,Deff}.

In this Letter, we focus on the fast loading protocol of a BEC into an optical lattice, 
which reproduces the result of adiabatic loading in a finite time.
The required driving potential is derived using the fast-forward theory  \cite{mas1,mas2,mas3}, and can be implemented 
using the time-averaging painting potential technique \cite{Miln,Frie,Hend}. Alternatively,
it can be efficiently approximated by  a time-dependent bichromatic optical lattice.

{\it Fast-forward loading scheme.--} 
The process of loading an atomic cloud into an optical lattice can 
 be described by a ramp, in a finite loading time $T_F$, of a periodic potential 
$\mathcal{V}[\q,R(t)]=R(t)V_F(\q)$,
where $V_F(\q)$ represents the optical lattice configuration at the end of the process, $t=T_F$, and $\q$ is the three-dimensional space coordinate.
Here, $R(t)$ is a switching function satisfying the boundary 
conditions $R(0)=0$ and $R(T_F)=1$.
Under such protocol the evolution is generally nonadiabatic.
Predominant excitations are originated in an early stage of the 
loading process \cite{Lack}, where the atomic cloud is weakly 
interacting and characterized by a large number of particles,
and a discretized lattice description is invalid.  
We pose the question as to whether there exists a 
loading scheme, which enforces the evolution
through the instantaneous ground state of the atomic cloud
in the potential $\mathcal{V}[\q,R(t)]$,
and seek an auxiliary potential $\mathcal{V_{\rm aux}}$ which assists the evolution.

We consider a BEC cloud described by the time-dependent
Gross-Pitaevskii equation (TDGPE),
\beqa
i\hbar\pa_t\Psi = -\frac{\hbar^2}{2m}\nab^2\Psi 
+ (\mathcal{V}+\hanaV_{\rm aux})\Psi + g |\Psi|^2
\Psi,
\label{TDGPE1}
\eeqa 
where $m$ is the mass of an atom, $g$ the coupling constant, and 
$\Psi$ the condensate wave function. 
In the fast-forward paradigm \cite{mas2} 
we look for an evolution parametrized by 
\beqa
\Psi(\q,t)=\psi[\q,R(t)]e^{i\phi(\q,t)}e^{-\frac{i}{\hbar}\int_0^{t} \mu[R(t')] dt'}.
\label{Psi1}
\eeqa
where 
$\phi(\q,t)$ is the condensate phase,
$\mu(R)$ the chemical potential of the instantaneous ground state and
$\psi[\q,R(t)]$
the mean-field condensate wave function of the ground state,
which satisfies 
\beqa
-\frac{\hbar^2}{2m}\nab^2\psi 
+ \mathcal{V}\psi + g |\psi|^2
\psi = \mu \psi.
\label{GPE1}
\eeqa
We impose the boundary condition $\phi(\q,0)=\phi(\q,T_F)=0$, so that
the initial and the final states coincide with the ground states
$\psi[\q,R(0)]$ and 
$\psi[\q,R(T_F)]e^{-\frac{i}{\hbar}\int_0^{T_{F}} \mu[R(t')] dt'}$ at $t=0$ and $T_F$,
respectively.
We divide both sides of TDGPE (\ref{TDGPE1}) 
by $\Psi$ and substitute Eq. (\ref{Psi1}) and decompose the equation into 
the real and imaginary parts.
We then use Eq. (\ref{GPE1}) and the fact that $\psi[\q,R(t)]$ 
can be taken to be real. The real part leads to the 
form of $\mathcal{V}_{\rm aux}(\q,t)$ as \cite{mas2}
\beqa
\mathcal{V}_{\rm aux}(\q,t)&=&-\frac{\hbar^2}{2m}(\nabla\phi)^2
-\hbar  \pa_t \phi,
\label{FFpot2}
\eeqa
and the imaginary part leads to 
\beqa
\nabla^2\phi+2\nabla\ln\psi\cdot\nabla\phi+\frac{2m}{\hbar}\dot{R}\partial_R\ln\psi=0,
\label{cc2}
\eeqa
which is used to obtain the phase $\phi$.
Equipped with both $\psi[\q,R(t)]$ and $\phi(\q,t)$
obtained by solving Eqs. (\ref{GPE1}) and 
(\ref{cc2}), the auxiliary
potential can be directly obtained from Eq. (\ref{FFpot2}).
By choosing $\dot{R}=\ddot{R}=0$ at $t=\{0,T_F\}$, 
$\phi(\q)$ and the auxiliary potential 
can be taken to be zero at $t=\{0,T_F\}$ [see Eqs. 
(\ref{FFpot2}) and (\ref{cc2})].
The composite ``fast-forward'' driving potential 
$\mathcal{V}_{\rm FF}=\hanaV + \hanaV_{\rm aux}$
generates $\psi[\q,R(T_F)]e^{-\frac{i}{\hbar}\int_0^{T_{F}} \mu[R(t')] dt'}$ from $\psi[\q,R(0)]$ in time $T_F$.

{\it Numerical results.--}
We consider a  one-dimensional model describing 
a cigar shaped BEC in an elongated 
anisotropic trap, 
and parametrize the time dependence of $R$ by
\begin{eqnarray}
R(t) = \frac{1}{T_F} \Big{[} t - \frac{T_F}{2\pi}\sin \Big{(}\frac{2\pi}{T_F}t\Big{)}
\Big{]}.
\label{eqR}
\end{eqnarray} 
Notice that this ramp function 
satisfies the boundary conditions $R(0)=0$, $R(T_F)=1$ and that its first  
and the second
derivatives vanish 
at $t=\{0,T_F\}$.
The optical lattice potential is taken to be
\begin{eqnarray}
\hanaV[q,R(t)] = U_F R(t)\sin^2 (k_Lq), 
\label{hanaV2}
\end{eqnarray}
where $q$, $L=\pi/k_L$ and $U_F$ are the
one-dimensional space coordinate along the longitudinal axis of the BEC,
the period of the lattice potential and the 
height of the lattice potential at $T_F$, respectively.

\begin{figure}[t]
\begin{center}
\includegraphics[width=0.65\linewidth]{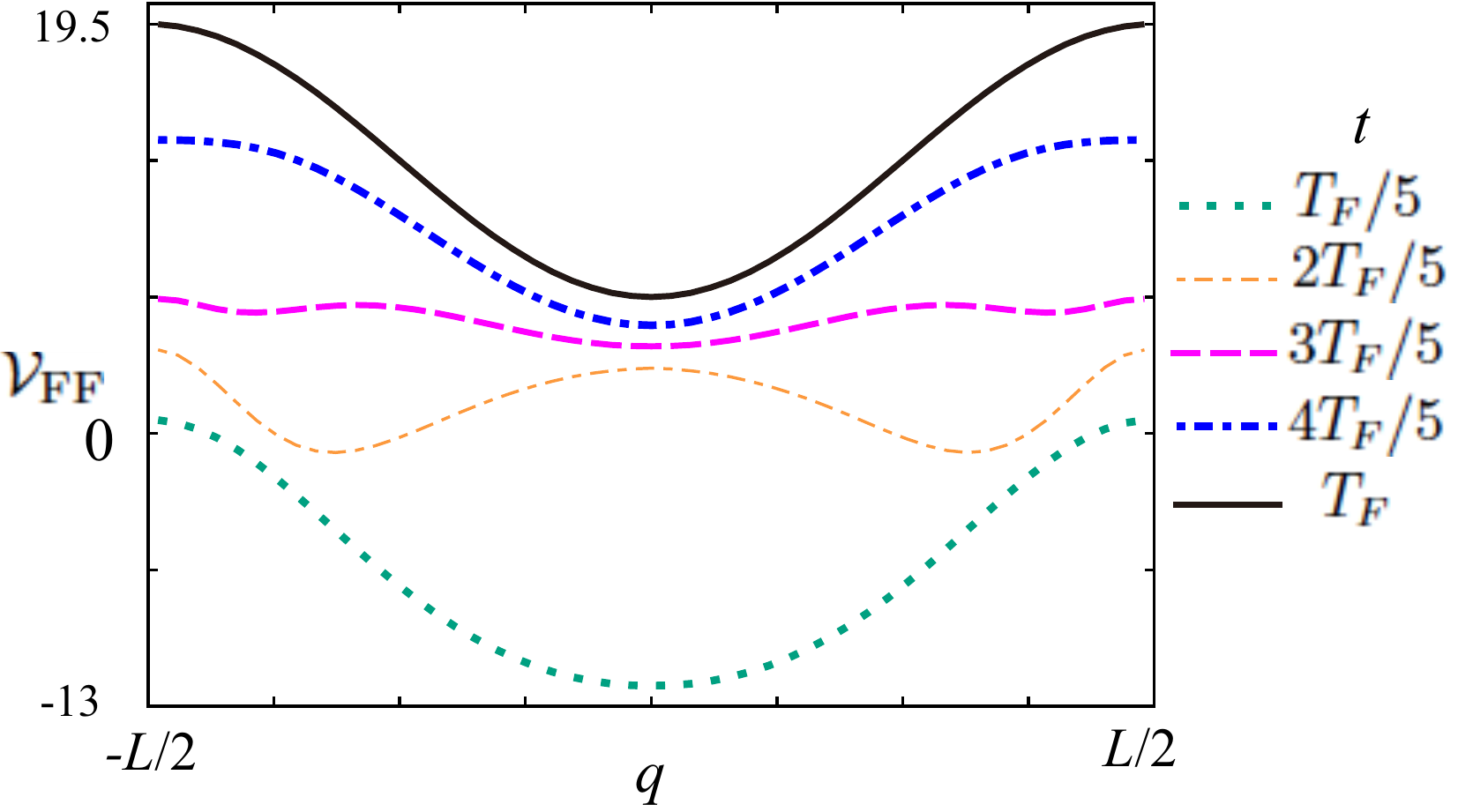}
\end{center}
\caption{\label{fig:epsart} Single-site driving potential 
$\hanaV_{\rm FF}(q,t)$ during the fast loading of an optical lattice in units of $E_R$ for several fractions of the loading time $T_F$. 
$\hanaV_{\rm FF}(q,t)$ for $t>T_F/5$ is shifted upward for the comparison.
}
\label{VFF2}
\end{figure}

The driving potential is a functional of $\psi$.
We first assume a homogeneous scenario with a spatially uniform BEC as
initial state (the effect of an external longitudinal harmonic trap will be 
discussed below).
For a precise characterization of the breakdown of adiabaticity in the 
homogeneous case we shall focus on the single-site dynamics of the
condensate wave function governed by TDGPE with effective one-dimensional 
coupling constant $c=gN/N_{\rm well}$,
where $N_{\rm well}/N$ is the fraction of the total number of atoms in a
single well.
We use the mass of $^{87}$Rb for $m$ and 
choose $T_F=27.4\ \mu$s, $L=0.4$ $\mu$m
and $U_F = 13E_R$ in terms of the recoil energy $E_R = (\hbar k_L)^2/2m$,
for which a loading scheme with $\mathcal{V_{\rm aux}}=0$ leads to a highly nonadiabatic dynamics.
$\psi(q,R)$ is obtained  numerically  by finding the instantaneous ground states
of Gross-Pitaevskii equation as the optical lattice is ramped up.
Engineering a STA is  found to be most difficult in the
non-interacting case ($c=0$), that we discuss first.
The driving potential $\hanaV_{\rm FF}(q,t)$
is shown in Fig. \ref{VFF2} for various times for $c=0$.

\begin{figure}[t]
\begin{center}
\includegraphics[width=0.55\linewidth]{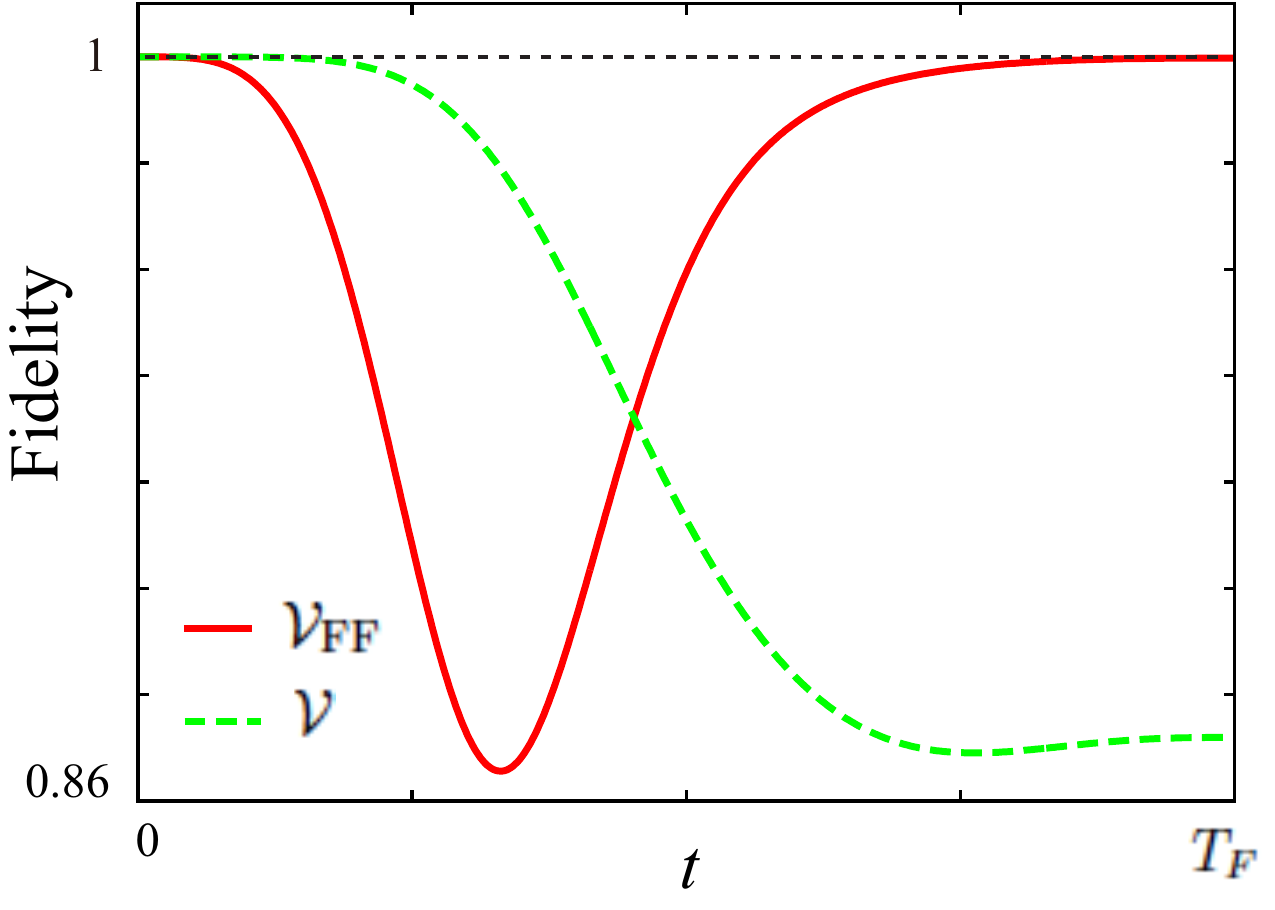}
\end{center}
\caption{\label{fig:epsart} Time-dependence of the fidelity between the instantaneous 
ground state of the optical lattice and the state resulting from a nonadiabatic driving. 
The final time is chosen such that a straightforward loading scheme based on $\hanaV$ fails (green dashed line),
illustrating the power of the fast-forward protocol under 
$\hanaV_{\rm FF}(q,t)$, which achieves unit-fidelity at the end of the process, $t=T_F$.
The nonadiabatic evolution under 
$\hanaV_{\rm FF}(q,t)$ is manifested in the transient low-fidelity for intermediate times.
}
\label{fid_com}
\end{figure}

\begin{figure}[t]
\begin{center}
\includegraphics[width=1\linewidth]{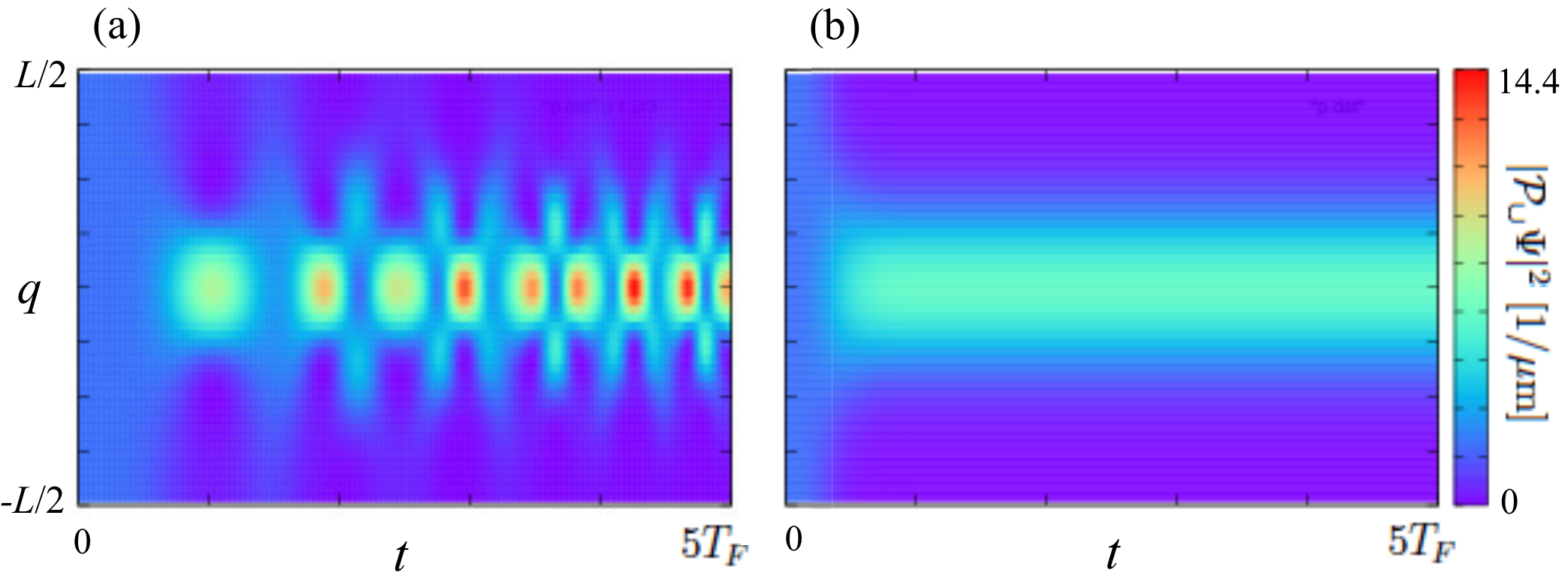}
\end{center}
\caption{\label{fig:epsart} Shortcut to adiabaticity during the fast loading of an infinite optical lattice. 
The time-evolutions of the density profile in a single-site are compared between the dynamics
driven by 
 (a)  $\hanaV[q,R(t)]$
 and (b) $\hanaV_{\rm FF}(q,t)$.}
\label{p}
\end{figure}

The evolution of $\Psi(t)$ is monitored by the single-site fidelity
$F(t) = |\la P_\cup\psi[R(t)]|P_\cup\Psi(t)\ra|$,
where $P_\cup$ is the single-site projector,
see Fig. \ref{fid_com}.
Whenever the optical lattice is ramped following the direct protocol 
$\hanaV[q,R(t)]$
in Eq. (\ref{hanaV2}), the breakdown of adiabatic dynamics 
leads to a substantial decay of the final fidelity
(see the dashed line in Fig. \ref{fid_com}).
We shall term the control $\hanaV[q,R(t)]$ as simple control hereafter.
By contrast, whenever the fast-forward protocol is used, the 
evolution is nonadiabatic along the process, 
and the  phase modulation $\phi(q,t)$ of the time evolving state  is responsible for the 
decrease of the fidelity at intermediate times along the process, 
i.e. $0<t<T_F$, as shown by the solid curve in Fig. \ref{fid_com}.
Nonetheless,  the fidelity becomes unity again at the final time,
as the target state is reached exactly.
The time-evolution of the density profile $|P_\cup \Psi|^2$
under $\hanaV[q,R(t)]$ and 
$\hanaV_{\rm FF}(q,t)$
is exhibited for $0\le t\le 5T_F$ 
in Figs. \ref{p}(a) and \ref{p}(b), respectively.
In both controls, the potential is fixed to $V_F$ for $t>T_F$.
The simple control excites the single-site breathing as manifested by the
temporal oscillation of $|P_\cup\Psi|^2$ 
in Fig. \ref{p}(a).
By contrast, these excitations are completely suppressed under the
$\hanaV_{\rm FF}$ driving, which induces
a nonadiabatic loading of the BEC in the ground state of the final optical 
lattice as appreciated in the evolution of the density profile
$|P_\cup\Psi|^2$ in Fig. \ref{p}(b), which becomes 
stationary for $t>T_F$.

\begin{figure}[t]
\begin{center}
\includegraphics[width=0.7\linewidth]{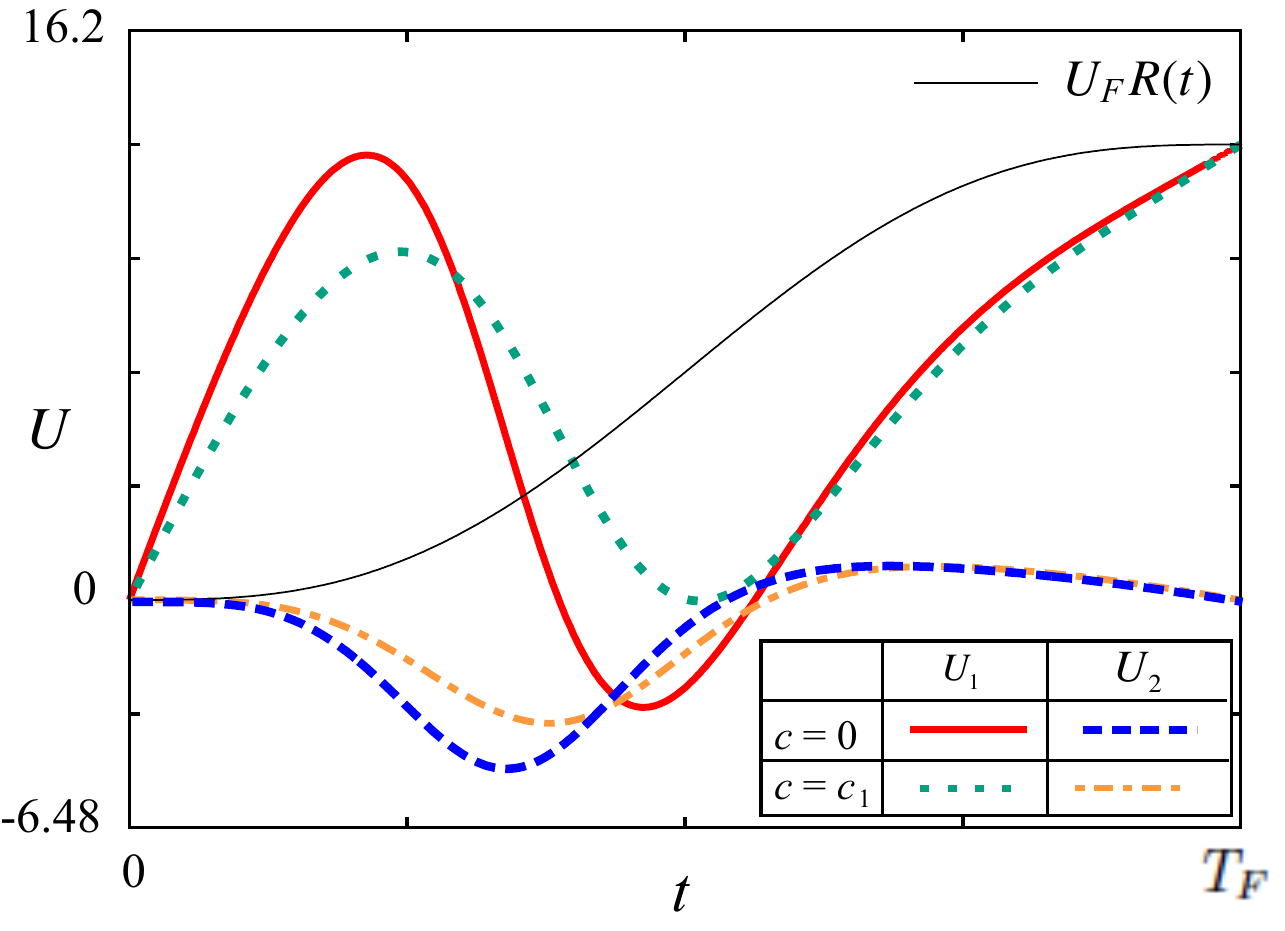}
\end{center}
\caption{\label{fig:epsart}  Time-dependence of $U_j$ is shown 
with unit $E_R$ for $c=0$ and $c_1$.
The thin black solid curve represents the time dependence of 
$U_FR(t)$ of Eq. (\ref{hanaV2}) in the simple control. 
}
\label{a_b_paper_T1_0}
\end{figure}

{\it Fast loading with a bichromatic optical lattice. -- }
The driving potential shown in Fig. \ref{VFF2} can be
implemented using, for instance, the painting technique,
which uses a rapidly moving laser beam to create a possibly dynamic
time-averaged optical dipole potential \cite{Miln,Frie,Hend}.  
In what follows,  
we propose an alternative and broadly accessible protocol, which resorts
instead to the use of a composite bichromatic optical lattice potential
of the form
\begin{eqnarray}
\mathcal{V}_{\rm app}(q,t) = U_1(t) \sin^2 (k_Lq)
+ U_2(t) \sin^2 (2k_Lq).
\label{Vapp}
\end{eqnarray}
The amplitudes $U_j(t)$ ($j=1,2$) are designed
using the least squares method
so that the composite lattice potential 
approximates the exact driving potential $\hanaV_{\rm FF}$.
The time dependence of $U_j$ shown in Fig. \ref{a_b_paper_T1_0} 
is calculated for $c/\hbar=0$ and $c/\hbar= c_1/\hbar=7.33$ mm/s with the same 
parameters as Fig. \ref{VFF2}.
The composite lattice driving potential in Eq. (\ref{Vapp}) 
succeeds in preparing the ground state
with high fidelity,
$F(T_F) = 0.9990$ and $0.9998$ for $c=0$ and $c_1$,
respectively, illustrating the fact that repulsive contact interactions
enhance the efficiency of the protocol by suppressing the formation of 
density ripples and other quantum transients in the cloud \cite{Camp142}.

We next impose a further constraint, and consider the case: 
$U_1\ge 0$ and $U_2\le 0$ 
to avoid quick phase shifts of the laser fields
required to induce an effective change in sign of $U_j$.
We proceed as above to find second approximation to the fast-forward driving
potential, which we denote by $\mathcal{V}_{\rm app}'$
with amplitudes $\{U_j'\}$.
$U_1'= 0$ when $U_1 <0$, and $U_2'=0$ when $U_2 >0$.
The time dependence of the fidelity for the control with 
$\mathcal{V}_{\rm app}'$
is almost the same as the one for  $\mathcal{V}_{\rm app}$.
The loading with $\mathcal{V}_{\rm app}'$ gives $F(T_F)=0.998$ and $0.9997$ 
for $c=0$ and $c_1$, respectively.
The range of $U_j$ for $c=c_1$ is narrower than that for $c=0$
because the wave function flattens for $c>0$ and 
its dynamics is more sluggish than in the non-interacting case.

\begin{figure}[t]
\begin{center}
\includegraphics[width=0.7\linewidth]{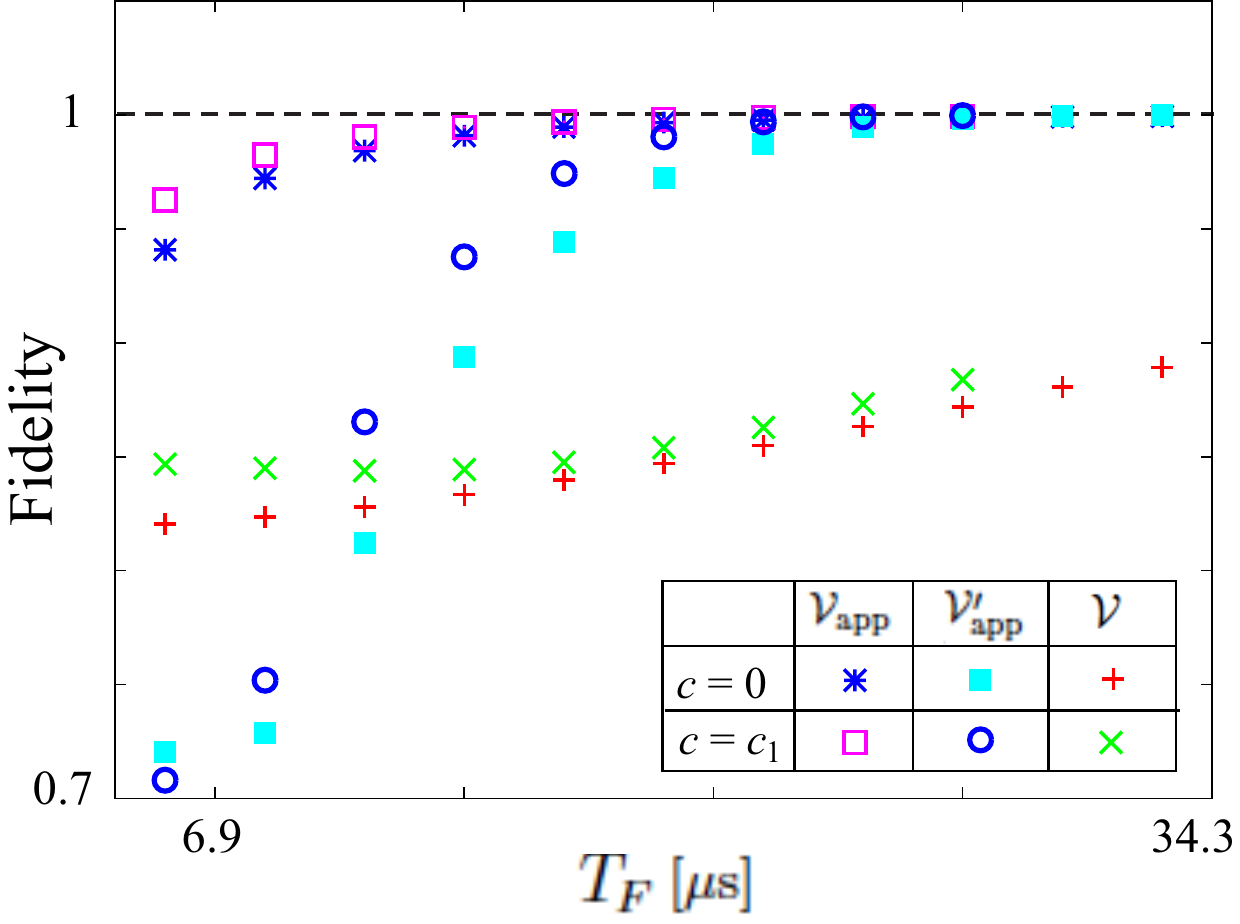}
\end{center}
\caption{\label{fig:epsart}  Efficiency
as a function of the loading time $T_F$ of a shortcut to the adiabatic loading of an optical lattice 
quantified by the fidelity $F(T_F)$ between the final ground state and the state resulting from several nonadiabatic loading schemes designed by the fast-forward technique. 
The dashed line represents $F=1$.}
\label{fid_com_ffpapp}
\end{figure}

In the control with the constraints,
$F(T_F)$ for $c>0$ is larger than the one for $c=0$ because
the time-interval, when $U_j'=0\ne U_j$, is shorter for
$c>0$ than that for $c=0$.
While the $\mathcal{V}_{\rm app}'$ control speeds up the preparation of the 
target state with high fidelity compared to the simple control for
this $T_F$,
the resulting fidelity can be degraded for smaller $T_F$ 
as the approximation of $\mathcal{V}_{\rm FF}$ by $\mathcal{V}_{\rm app}'$
worsens for small $T_F$. 
$F(T_F)$  as a function of $T_F$ is shown in Fig. \ref{fid_com_ffpapp}.
The loading under $\mathcal{V}_{\rm app}$ is effective 
even for $T_F=5.5\ \mu$s, while $\mathcal{V}_{\rm app}'$
disturbs the BEC more than the simple control $\mathcal{V}$ for such a short loading time for the above reason.
In addition, $F(T_F)$ for $c=c_1$ is larger than that for $c=0$
because the bichromatic lattice potential 
mimics the exact driving potential more closely in the interacting case.

\begin{figure}[t]
\begin{center}
\includegraphics[width=0.9\linewidth]{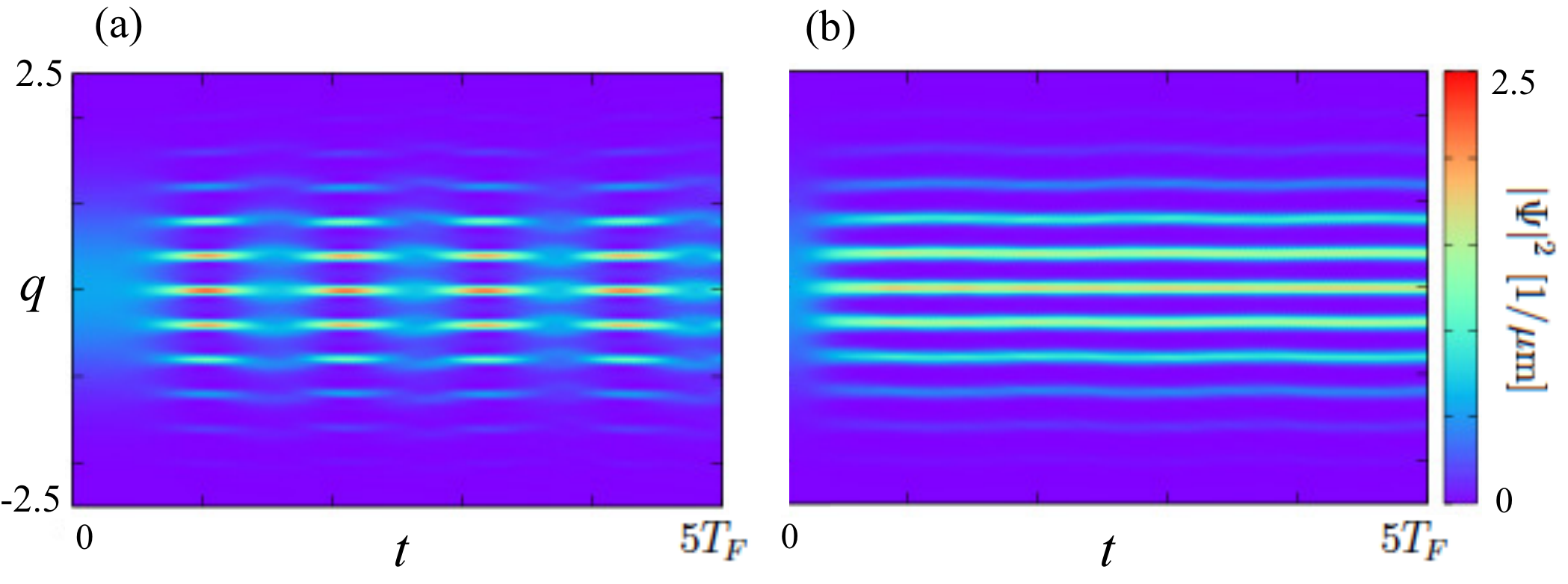}
\end{center}
\caption{\label{fig:epsart}  Time evolution of the density profile during the fast loading of an optical lattice in the presence of an external  harmonic confinement. 
(a) The straightforward loading scheme driven by $\hanaV[q,R(t)]$ leads to a breakdown of adiabaticity resulting in the excitation of the on-site breathing mode.
(b) Whenever the nonadiabatic protocol resorts in a time-dependent bichromatic lattice approximating the fast-forward potential $\hanaV_{\rm app}'(q,t)$, the dynamics of fast loading mimics the adiabatic protocol and excitations are suppressed. }
\label{p_large_c0}
\end{figure}

So far we have neglected the effect of the longitudinal confinement
whose main effect is to introduce a local chemical potential.
In what follows we study the efficiency 
of our method in the presence of an external harmonic trap
by applying $\hanaV_{\rm app}'$ to the ground state.
We consider two cases: 
(a) $\omega=2\pi\times 116$ Hz with $T_F=27.4\ \mu$s
and (b) $2\pi\times 1.8$ Hz with $T_F=13.7\ \mu$s 
for $c=0$ with the same parameters
as Fig. \ref{VFF2}.
The wave function of the target ground state extends over 
7 and 60 wells of the lattice potential, respectively. 
The fidelity at $T_F$ is $0.909$ in case (a) and $0.910$ in case (b),
while in the simple control $F(T_F)=0.795$ and $0.67$, 
respectively.
In Fig. \ref{p_large_c0},
time-evolutions of $|\Psi|^2$ under $\hanaV[q,R(t)]$ and $\hanaV_{\rm app}'$
are exhibited for $0\le t\le 5T_F$ for case (a).
In the both controls, the lattice potential is fixed to $V_F$ for $t>T_F$. 
In the simple control, excitations manifest in
the time evolution of the density profile $|\Psi|^2$ 
as a violent beating pattern, 
which can be suppressed using the $\hanaV_{\rm app}'$
driving.
We have also studied the efficiency of our control in the case with
finite coupling constant $gN/\hbar=36.7$ 
mm/s with the harmonic confinement with $\omega=2\pi\times 464$ Hz 
for $T_F=27.4\ \mu$s.
Because of the local chemical potential induced by the harmonic confinement,
the fast-forward driving protocol is site-dependent.
However, we show that the driving potential $\hanaV_{\rm app}'$
designed for the average mean-field interaction over the occupied sites in
the final configuration remains efficient even in highly inhomogeneous systems.
To illustrate this we consider the case in which the 
final state extends over five lattice sites and
take $c(=gN/N_{\rm well})=c_1=7.33$ $\hbar$ mm/s as well as the same $U_j'$ used
in Fig. \ref{a_b_paper_T1_0}.
The fidelity at $t=T_F$ is around 0.972 in the control with $\hanaV_{\rm app}'$,
while $F(T_F)=0.865$ in the simple control.

{\it Conclusions. --}
A high-fidelity nonadiabatic loading scheme of a 
Bose-Einstein condensate into an
optical lattice has been engineered using the fast-forward approach.
The finite-time driving protocol generates the ground state in 
the final optical lattice potential starting from either a uniform or 
trapped atomic cloud, without unwanted residual excitations associated with
the breakdown of adiabaticity.
The required auxiliary potential supplementing the ramp of the optical
lattice can be implemented using the painting potential technique
\cite{Miln,Frie,Hend}.
We have further shown that a composite time-dependent bichromatic
lattice mimics the fast-forward driving potential with high fidelity.
Before closing, we point out that our approach can be further 
applied to vibrational multiplexing in optical lattices targeting the
preparation of excited states, extending recent proposals for harmonic
traps \cite{Mart14}. Equivalent control schemes can be designed as well 
in other systems, e.g., ultracold fermions in tight waveguides \cite{Kolomeisky00,Girardeau06}.

\begin{acknowledgments}
It is a pleasure to acknowledge discussions with V. Ahufinger, R. Onofrio, B. Rauer and L. Tarruell.
SM thanks Grants-in-Aid for Centric Research of 
Japan Society for Promotion of Science
and JSPS Postdoctoral Fellowships for Research Abroad
for its financial support.
This research is further supported by the U.S Department of Energy through the LANL/LDRD Program and a  LANL J. Robert Oppenheimer Fellowship (AD).

\end{acknowledgments}

\end{document}